\nonstopmode
\documentclass{aa}
\usepackage{graphicx}
\usepackage{lscape}

\begin{document}
\def\lsim{\, \lower2truept\hbox{${<\atop\hbox{\raise4truept\hbox{$\sim$}}}$}\,}
\def\gsim{\, \lower2truept\hbox{${>\atop\hbox{\raise4truept\hbox{$\sim$}}}$}\,}
   \title{High-frequency polarization properties of southern K\"uhr
sources}


   \author{R. Ricci\inst{1,2}
          \and
          I. Prandoni\inst{3}
          \and
          C. Gruppioni\inst{4}
          \and R.J. Sault\inst{2}
          \and G. De Zotti\inst{5,1}
          }

   \offprints{R. Ricci}

   \institute{SISSA/ISAS, Via Beirut 2--4, I-34014 Trieste, Italy \\
\email{ricci@sissa.it}
         \and
Australia Telescope National Facility, CSIRO, P.O. Box 76, Epping,
    NSW 2121, Australia \\
\and Istituto di Radioastronomia, CNR, Via Gobetti 101,
I--40129, Bologna, Italy \\
\and INAF, Osservatorio Astronomico di Bologna, Via Ranzani 1,
I--40126 Bologna, Italy \\
\and INAF, Osservatorio Astronomico di Padova, Vicolo
dell'Osservatorio 5, I--35122 Padova, Italy \\
\email{dezotti@pd.astro.it}
             }

   \date{Received ...... ; accepted .......}

   \abstract{We have observed 250 of the 258 southern sources in the
   complete 5 GHz 1 Jy sample by K\"uhr et al. (1981)
   using the Australia Telescope Compact Array (ATCA) at 18.5~GHz.
   This paper focuses on the polarization properties of this sample,
   while other properties
   will be addressed in a future paper.
   The analysis subdivides the sample into flat and steep spectrum
   sources using the classification of Stickel et al. (1994), 
   where spectral indices were measured between 2.7 and 5 GHz.
   The polarized flux has been measured with a $S/N > 5$ for 170
   sources (114 flat-spectrum and 56 steep-spectrum). Upper
   limits have been set for additional 27 sources (12 flat-spectrum and 15
   steep-spectrum). The median fractional
   polarization at 18.5 GHz for the flat-spectrum
   sub-sample is $\Pi_{18.5}\simeq 2.7\%$, which is
   about a factor of 2 higher than at 1.4
   GHz ($\Pi_{1.4}\simeq 1.4\%$, based on NVSS data).
   For flat-spectrum sources we find a weak correlation between
   $\Pi_{18.5}$ and the high-frequency (5--18.5 GHz) spectral index.
   No evidence was found
   of significant correlations of the fractional polarization with other source
   properties. The median value of $\Pi_{18.5}$ for the
   steep spectrum sources is $\simeq 4.8\%$, but our
   sample might be biased against extended sources.
   We find some correlation between $\Pi_{18.5}$ and both
   the low-frequency (1.4--5 GHz) and the high-frequency 
   (5--18.5 GHz) spectral indices.
   An important application of this work is in estimating
   the contamination of CMB polarization maps by extragalactic radio sources.
   Our results indicate that such contamination
   is within that estimated by Mesa et al. (2002).

   }

   \maketitle
%

\section{Introduction}

Measurements of polarization properties of extragalactic radio
sources provide unique information on magnetic field
structure. The fractional polarization is a measure of the orderedness
of the
magnetic field in the emitting region. If synchrotron
self-absorption and Faraday rotation can be neglected,
the position angle will be
orthogonal to the magnetic field orientation.
Given this, it is
surprising that high radio frequency polarimetric surveys
of {\it complete} samples of extragalactic radio sources are so
rare.

The general high-frequency polarization properties
of radio sources are poorly known. Multifrequency observations
by Jones et al. (1985) and Rudnick et al. (1985) of 20
flat-spectrum  sources ($\alpha < 0.5$; $S \propto \nu^{-\alpha}$)
showed that the median value of polarization is $\sim 2.5\%$ in the range
1.4--90 GHz, independent of frequency.  This would suggest 
that the general polarization properties can be accounted for by
largely random magnetic fields. Note, however, that
Jones et al. and Rudnick et al. stressed that
their sample is not complete and selection
criteria may introduce unpredictable biases.
On the other hand, the relatively constant polarization fraction
suggests that
differential Faraday rotation, or
decreased contribution of
relatively unpolarized opaque regions, or greater magnetic field
ordering on the smaller scales dominating at millimeter
wavelengths, may play minor roles (see also Saikia \&
Salter 1988). Okudaira
et al. (1993) made linear polarization observations at 10 GHz
for a complete sample of 93 flat-spectrum sources
selected at 5 GHz ($S_{5{\rm GHz}}> 0.8\,$Jy).
Okudaira et al. do not provide an average value for the fractional
polarization
at 10 GHz.
From their tabulated source values one could infer a median
fractional polarization of 2.3\%. On the other hand,
Aller et al. (1999) find that the fractional linear polarization
of a incomplete sample of 41 BL Lac objects is generally
higher at 14.5 GHz (median value 5\%) than at 4.8 GHz (median
value 3.6\%). Also, the millimeter/sub-millimeter polarization
survey of 26 blazars by Nartallo et al. (1998) yielded a median
fractional polarization at 0.8--1.1 mm of 7.1\%, while the NVSS data
indicate, for the same objects, a median fractional polarization of
$\simeq 2\%$ at 1.4 GHz (Mesa et al. 2002). For a sample of
steep-spectrum sources with multifrequency polarimetric
observations by Klein et al. (2003), Mesa et al. (2002) find a
steady increase with frequency of the mean fractional polarization
from $\simeq 3\%$ at 1.4 GHz to 8.65\% at 10.6 GHz. It should be
noted, however, that the sample is biased towards high
polarizations as it includes
only sources which have been detected in polarization 
at 10.6 GHz.
The same effect may occur in the high redshift radio galaxy
sample observed by Pentericci et al. (2000) at high resolution (0''.2),
where typical fractional polarization of 10--20\% at 8.4 GHz were found.

A good understanding of the high-frequency polarization
properties of extragalactic sources is crucial in analysing
 the polarization of the cosmic microwave background (CMB).  This has
been detected recently by the DASI experiment at 30~GHz (Kovac et al. 
2002) and by the WMAP satellite (Kogut et al.  2003) in several bands in
the range 23--94 GHz.  The extraction of the very weak CMB polarization
signal requires both high sensitivity and careful control of
foregrounds.  In the frequency range relevant to these experiments, the
polarized emission of extragalactic radio sources is expected to be the
main contaminant on small angular scales (Sazhin \& Korol\"ev 1985; De
Zotti et al. 1999). 

In this paper we describe the results of linear polarization
measurements at 18.5 GHz of 250 of the 258 southern
($\delta < 0^\circ$) extragalactic sources in the 5 GHz K\"uhr et
al. (1981) all-sky sample. This sample is complete to 
$S_{5{\rm GHz}} \ge 1\,$Jy,
and contains 527 objects. This is, by far, the
largest {\it complete} sample of radio sources for which
high-frequency polarization measurements have been carried out.
Observations and data reduction are described in Sect. 2 and 3,
respectively. The results are analyzed and discussed in Sect. 4. In
Sect. 5 we summarize our main conclusions.

\section{Observations}

Measurements were carried out with the Australia Telescope
Compact Array (ATCA). Since 2001 this facility has been undergoing
a major millimeter upgrade that will provide 12 mm and 3 mm
receivers on all the ATCA antennas. In the commissioning phase,
prototype 12 mm receivers were mounted on antennas
CA02, CA03 and CA04. We took snapshot observations of the southern
sources in the so-called EW367 array
configuration.
The two frequency bands were centered at 18.5 and 18.65 GHz with each
band having
a bandwidth of 128 MHz. The array configuration that we used
is quite compact: the longest baseline was 214.3~meters, giving
a beamwidth of 15.6 arcsec.

The observations were performed in three runs during March 2002.
The total observing time was 48 hours: 24 hours on March 20-21, 10 hours on
March 21-22 and 14 hours on March 27-28.
We used the ``mosaic observing mode''
because this allowed us to organize the observing sequence to
minimize the slew time between the sources. The 258 sources were
split in 36 clusters of sources. Clusters were observed
four times each to produce a good spread in hour angles.

Many of these sources had already been targeted with the ATCA
as part of a search for potential 22 GHz calibrators. The 22 GHz total
intensity measurements with ATCA of the full sample were completed
in January 2001; the results will be reported in a subsequent
paper.  These measurements were used to determine the integration
time for each target source to ensure a detection at $\geq 5\sigma$
for source polarizations $\geq 1\%$.
Nevertheless, because of limited observing time,
we set an upper limit of
30 minutes per source. This means that for 43 sources with 22 GHz
flux density $\leq$ 0.192 Jy, the detection threshold is $\geq$ 1\%.
Because of technical problems during the runs,
8 sources were not successfully observed.

\section{Data reduction}

The radio source 1934-638 was used as the primary flux density
calibrator, while phase and gain calibrators were chosen
from the ATCA calibrator list.
The instrumental polarization was calibrated using
measurements of Jupiter, which we assumed to be unpolarized.

Unfortunately the atmospheric phase stability 
during the observations was poor, so it was not 
possible to image the sources. 
Instead we decided to use
non-imaging model-fitting techniques.
To measure calibrators sufficiently 
frequently to correct for 
phase instability would have been quite expensive in time:
this would have reduced
significantly the number of sources that we could 
observe in the scheduled time.

The data were reduced in the MIRIAD
software package (Sault et al., 1995) to derive
source flux densities in all the Stokes parameters (I,Q,U,V).
In particular we
used the MIRIAD task CALRED to determine the
flux densities. In processing the visibilities, CALRED assumes a point source
model, and produces a flux density estimate for
each Stokes parameter for {\it each scan}. Along with the
flux density and its error, the task computes out the theoretical rms noise,
and a confusion parameter, $C$. The confusion parameter is
an estimate of the fraction of the flux density that fails to conform to a 
point-source model.

\begin{figure}
   \centering
\includegraphics{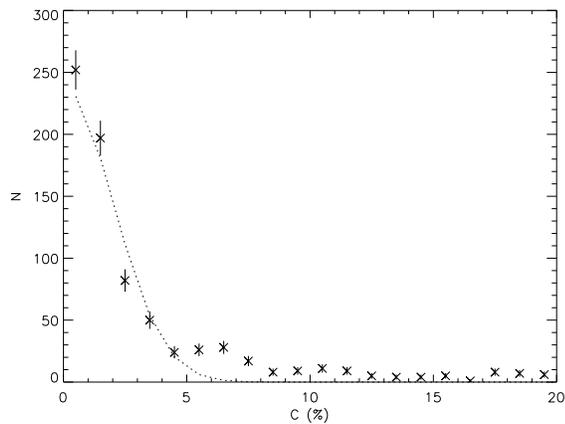} \vspace{6 cm}
\caption{Distribution of the confusion
parameter $C$. A semi--Gaussian fit is also shown (dotted line).
The fitted standard deviation is $\sigma$=2.03.}
\label{fig:confhisto}
\end{figure}

A high value of $C$ means that the source is poorly modelled as a point
and that the determination of its flux
density is poor. On the other hand, even in the case of truly
point sources (where the value of $C$ should be 0), noise
results in a spread of the estimated value of $C$. For normal
error distribution, $C$ is approximately described by a semi-Gaussian 
distribution (negative values of $C$ are not allowed).
The observed distribution of
$C$ (Fig.~\ref{fig:confhisto}) shows a strong excess over the
best-fit Gaussian curve where $C \ge 5\%$: we believe
this is a result of
source extension.
Consequently we have chosen to exclude all the scans
with $C \ge 5\%$. This resulted in the rejection of 53 sources.


For the remaining 197 sources
 we computed the linearly polarized flux for each scan $i$,
$F_{p,i}=\sqrt{Q_i^{2}+U_i^{2}}$ and the associated error
$\sigma_{F,i}=\sigma_{Q,i}=\sigma_{U,i}$ ($\sigma_{Q,U}$ being the
rms noise on the Stokes parameter $Q$ or $U$). For all the sources
the polarized flux, $F_p$, and its error, $\sigma_F$, were obtained by
using a weighted mean over all the scans:

\begin{equation}
F_p=\sigma^2_F\sum_{i}\frac{F_{p,i}}{\sigma^{2}_{F,i}}
\end{equation}
\begin{equation}
\frac{1}{\sigma^{2}_F}=\sum_{i}\frac{1}{\sigma^{2}_{F,i}}
\end{equation}
For 170 sources the polarized flux was detected with
a signal-to-(theoretical)noise ratio $\geq5$.
We considered such values as reliable.
An upper limit of 5 $\times$ the theoretical noise was assigned to
the other 27  sources.
In all cases, the total intensity, $I$ (again derived as the weighted mean 
of the scan total intensities), was measured with very high $S/N$ ratio.
The observations did not determine the polarization 
position angle: the geometry on the ATCA 12mm receivers 
was not calibrated sufficiently well to give reliable position angle 
information.

In the following analysis we subdivide the sample into flat and steep-spectrum
sources. Following the classification of Stickel et al. (1994), we define as
flat the sources with spectral
index between 2.7 and 5 GHz $\alpha_{2.7}^5 < 0.5$
(S$_\nu \propto \nu^{-\alpha}$). Sources with $\alpha_{2.7}^5 > 0.5$ are
considered steep.
In table \ref{tab:fsss} we list the number (and the corresponding fraction
of the total) of detected sources for the
flat and steep-spectrum sub-samples. For both classes we also list
the upper limits and the rejected sources ($C$ $>$ 5\%).
While the final flat-spectrum source sample is almost
complete (88\%), only 49\% of the steep-spectrum sources have
reliable detections.

\begin{table}
\begin{center}
\caption{Source number statistics for the polarization observations.
The fractions in brackets are the percentages of flat and steep 
spectrum sources.}
\label{tab:fsss}
\begin{tabular}{|l|r|rr|rr|}
\hline
        &   All   & \multicolumn{2}{c|}{Flat}  &  \multicolumn{2}{c|}{Steep} \\
\hline
detections   & 170 & 114 & (80\%) & 56 & (49\%) \\
upper limits &  27 &  12 & (8\%) & 15  & (13\%)  \\
rejected     &  53 &  13 & (9\%) & 40 & (35\%)  \\
not observed &   8 &   4 & (3\%) &  4 & (3\%)   \\
\hline
total        & 258 &  143 & & 115 & \\
\hline
\end{tabular}
\end{center}
\end{table}

The total intensities and polarized fluxes for steep and
flat-spectrum sources are listed in Tables~\ref{table:SStot}
and~\ref{table:FStot}, respectively. Also listed are the linear
fractional polarization $\Pi_{18.5}=F_{p}/I$, the spectral
indices between 1.4 and 5 GHz ($\alpha_{1.4}^5$) and 5
and 18.5 GHz ($\alpha_{5}^{18.5}$) and the 5 GHz
luminosity. Source types and redshifts are from Stickel et al. (1994).

\section{Polarization properties of sources}

\subsection{Steep-spectrum sources}


\begin{figure}
   \centering
   \includegraphics[width=9cm]{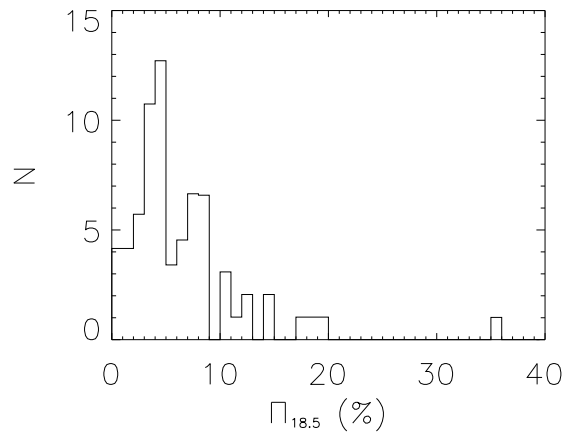}
\caption{Distribution of the fractional polarization at 18.5
GHz for the steep-spectrum sources.}
\label{fig:linpolSS}
\end{figure}

As expected, most of the rejected sources (40 out of 53 with $C>5\%$) are
steep-spectrum. This is because steep-spectrum sources are more
likely extended than flat-spectrum sources.
This means that the following analysis is based on an incomplete 
sample, which can be strongly biased against extended sources (see
table \ref{tab:fsss}).
Additionally only upper limits can be given for 15 out of 71 ($\sim$21\%) 
of the steep-spectrum sources.
Nevertheless some points are worth noticing. In Fig.\ref{fig:linpolSS}
we show the parent distribution of $\Pi_{18.5}$ for the steep-spectrum sample.
This was obtained by using the Kaplan--Meier estimator
[routine KMESTM in the software
package ASURV Rev 1.2 (Isobe \& Feigelson, 1990). This includes
upper limits by using
the survival analysis methods presented in Feigelson \& Nelson
(1985) and Isobe et al. (1986)]. The median of the
distribution is $\simeq 4.8\%$ and the mean is $\simeq 6.5\pm0.7\%$.
If we divide the sample into quasars (21 sources) and radio galaxies
(50 sources) we find median values of
4.4\% and 6.1\%, and mean values of 5.1$\pm$0.6\% and 7.0$\pm$0.9\%
respectively.
The Peto \& Prentice generalized Wilcoxon two sample test yields a
probability of 21.7\% that the distributions of QSOs and radio
galaxies are drawn from the same parent population.

These values are somewhat lower than
those found by Klein et al. (2003) at 10.6 GHz and by
Pentericci et al. (2000) at 8.4 GHz by  for their
sample of high-redshift radio galaxies observed at high angular
resolution (typical polarizations of 10--20\%).
Nevertheless we point out that both the high spatial resolution of
the Pentericci et al. sample and
the selection criterion used by Klein et al. may introduce a bias
toward high fractional polarization values, as stressed in the 
introduction.

\begin{table}
\begin{center}
\caption{Test for correlations between $\Pi_{18.5}$ and other
properties of the steep-spectrum sources. N is the number of
sources, Z is the generalized Kendall's correlation coefficient
and P is the probability that a correlation is not present. }
\label{tab:SScorr}
\begin{tabular}{|l|c|r|c|}
\hline
 Correlation   &   N   & \multicolumn{1}{c|}{Z}  &  P   \\
\hline
$\Pi_{18.5} - \alpha_5^{18.5}$    &  71  &  3.526  & 0.0004 \\
$\Pi_{18.5} - \alpha_{1.4}^5$     &  37  &  2.747  & 0.006 \\
$\Pi_{18.5} - z$                  &  49  &  1.283  & 0.1996  \\
$\Pi_{18.5} - \log L_{5\rm{GHz}}$ &  27  &  0.764  & 0.445 \\
$\Pi_{18.5} - \Pi_{1.4}$          &  37  &  0.526  & 0.5992 \\
\hline
\end{tabular}
\end{center}
\end{table}

We have used the NVSS public data (Condon et al. 1998) to
investigate the polarization properties of our sources at 1.4 GHz.
The 37 steep-spectrum sources detected by the NVSS in the
overlapping region, have a median 1.4 GHz fractional polarization of
$\simeq 0.4\%$, comparable to the residual instrumental
polarization (0.3\%). To obtain a more reliable estimate of the
1.4 GHz polarization, we included the
steep-spectrum K\"uhr sources in the Northern sky. This selected
184 objects in total. We find $\Pi_{1.4,{\rm median}}\simeq 1.1\%$.
The Peto \& Prentice two-sample test rules out at a very high
confidence level (P$<10^{-5}$) the hypothesis that the
distributions at 1.4 and 18.5 GHz are drawn from the same parent
population.

\begin{figure}
   \centering
\includegraphics{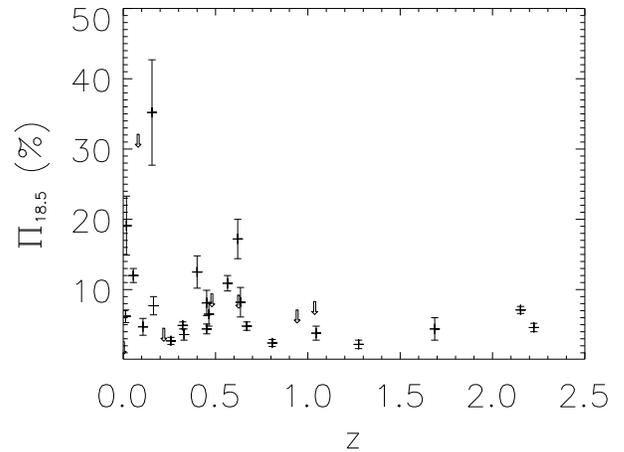} \vspace{7 cm} \caption{$\Pi_{18.5}$ vs
redshift for the steep-spectrum sample. $\Pi_{18.5}$
upper limits are represented by downward arrows.}
\label{fig:sszcorr}
\end{figure}

The origin of this frequency dependence is not well
understood. The multifrequency polarization measurements by Klein
et al. (2003), quoted by Mesa et al. (2002), indicated an increase
by a factor of $\simeq 3$ of the fractional polarization of
steep-spectrum sources between 1.4 and 10.4 GHz. This is
attributed to
Faraday depolarization in uniform slabs with effective rotation
measure $RM\simeq 260\,\hbox{rad}\,\hbox{m}^{-2}$. 
However, an increase of the fractional polarization
with frequency, as expected in the case of a
Faraday screen at the source, translates in a positive correlation
with redshift. For example, in the presence of a Faraday screen
located at redshift $z$, the observed rotation measure is related
to the intrinsic value by $RM_{\rm obs}=RM_{\rm intr}(1+z)^{-2}$.
However no significant correlation is found in our data
between $\Pi_{18.5}$ and
$z$ (see Fig~\ref{fig:sszcorr} and Table~\ref{tab:SScorr}).
This may be consistent with the
findings by Pentericci et al. (2000) that the fraction of powerful
radio galaxies with extreme Faraday rotation measures ($RM >
1000\,\hbox{rad}\,\hbox{m}^{-2}$) increases steeply with redshift.
This suggests that their average environment becomes increasingly
dense with increasing $z$. The larger mean fractional polarization
measured at 18.5 GHz may imply that Faraday depolarization is
negligible at high frequencies. Polarization
measurements at intermediate frequencies are necessary to
assess this.

\begin{figure}
   \centering
\includegraphics{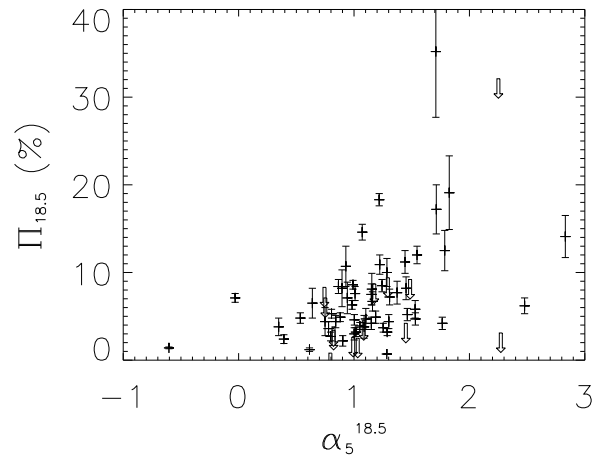} \vspace{7 cm} \caption{$\Pi_{18.5}$ vs
$\alpha_5^{18.5}$ for the steep-spectrum sample. $\Pi_{18.5}$
upper limits are represented by downward arrows.}
\label{fig:SScorr}
\end{figure}

Table~\ref{tab:SScorr} also shows the results of our
investigation of possible correlations of $\Pi_{18.5}$ with other
source properties. There are indications of a
correlation with the spectral indices $\alpha_{1.4}^5$ and
(more significantly) $\alpha_5^{18.5}$:
steeper sources seem to have stronger $\Pi_{18.5}$ (Fig.~\ref{fig:SScorr}).
The fact that sources with the highest $\Pi_{18.5}$ and the 
steepest $\alpha_5^{18.5}$ are generally at low redshift 
may suggest that they are resolved and we are actually seeing 
highly polarized portions of them, such as radio lobes.  
We note, however, that any correlation should be taken 
with caution due to the incompleteness of the steep-spectrum source sample.
The fractional polarization
at 18.5 and 1.4 GHz are uncorrelated. We do not
see any correlation between fractional polarization and other
source properties.

\begin{figure}
   \centering
   \includegraphics[width=9cm]{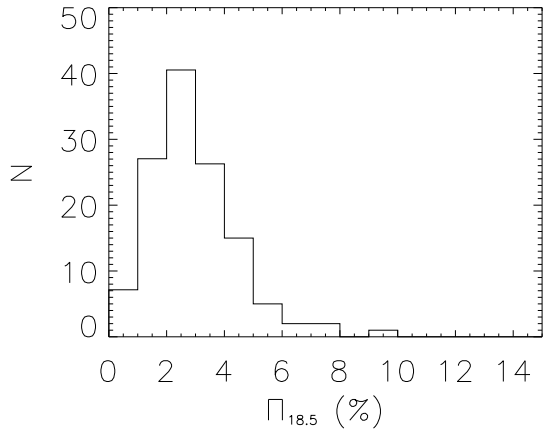}
\caption{Distribution of the fractional polarization at 18.5
GHz for the flat-spectrum sample.}
\label{fig:linpolFS}
\end{figure}

\begin{figure}
   \centering
   \includegraphics[width=9cm]{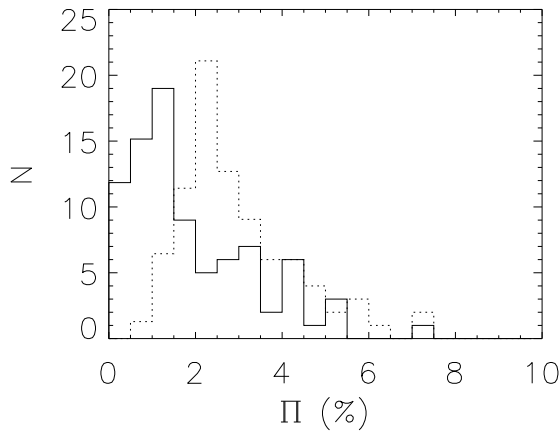}
   \caption{Linear fractional polarization
   distribution at 1.4 GHz (solid line) and 18.5 GHz (dotted line)
   for the 86 flat-spectrum southern K\"uhr sources in the NVSS survey.}
   \label{fig:linpolNVSS}
\end{figure}

\subsection{Flat-spectrum sources}

We used the Kaplan-Meyer estimator to derive the fractional
polarization distribution of the 126 flat-spectrum sources
in our final sample (114 detections and 12 upper limits). 
The result is shown in
Fig.~\ref{fig:linpolFS}: the median value is 2.7\% and the mean value
is 2.9$\pm0.1$\%. This is about a factor of 2 lower than that found for the
steep-spectrum sources.

We used the NVSS
catalog (Condon et al. 1998) to determine
the fractional polarization at 1.4 GHz, $\Pi_{1.4}$:
86 flat-spectrum sources overlap with the NVSS survey. The distribution of
$\Pi_{1.4}$ is shown in Fig.~\ref{fig:linpolNVSS}: it has a median
value of $1.4\%$, comparable to the value found for steep-spectrum sources,
and a factor of 2 lower than that at 18.5 GHz. This seems to be in disagreement
with the results of Jones et al. (1985) and Rudnick et al. (1985), who
found a constant polarization value of 2.5\% in the frequency range 
1.4--90 GHz.

In Table~\ref{table:medianFS} we give the median fractional polarization
for different classes of objects in the flat-spectrum sample
(QSOs, BL Lacs and radio galaxies). In the last column of the table, we
give the median fractional polarization at 1.4 GHz that was
computed by including the
northern ($\delta > 0$) K\"uhr sources with NVSS detection.
Adding the latter sources
increases the total number of flat-spectrum sources with 1.4 GHz
polarization data from 86 to 238: the number of QSOs increases
from 74 to 175, that of BL Lacs from 5 to 34, and that of galaxies
from 5 to 25.
We do not find any significant indication of differences in the 18.5 GHz
polarization properties of the various sub-populations. However
the very small number of BL Lacs and radio galaxies means this
comparison is not very meaningful.
The comparison at 1.4 GHz is more significant: the Peto \& Prentice 
generalized Wilcoxon two sample test
yields probabilities of $\simeq 1\%$ and of $0.001\%$ that the
fractional polarization distributions of QSOs are drawn from the same
parent population as BL Lacs and radio galaxies, respectively.

\begin{table}
\begin{center}
\caption{Median fractional polarization of flat-spectrum sources.
Columns 2 and 5 give the number and the median fractional polarization
at 18.5 GHz of the different classes of sources in our sample.
In parentheses are the number of such sources in the NVSS area
(column 3), and their median fractional polarization at 18.5 GHz
(column 6) and at 1.4 GHz (column 7). Columns 4 and 8 give
the total number of K\"uhr flat-spectrum sources of
the different classes in the NVSS catalog, and their median
fractional polarization at 1.4 GHz.} \label{table:medianFS}
\begin{tabular}{|l|rrr|rr|rr|} \hline
 Type       &\multicolumn{3}{c|}{N}&
 \multicolumn{2}{c|}{$\Pi_{18.5}\,(\%)$}
 &\multicolumn{2}{c|}{$\Pi_{1.4}\,(\%)$}\\
\hline
QSO         & 108 & (74) & 175 & 2.7 & (2.6) & (1.4) & 1.4 \\
BL Lac      &   8 &  (5) &  34 & 3.0 & (2.9) & (1.2) & 2.2 \\
GAL         &   8 &  (5) &  25 & 2.4 & (2.5) & (1.8) & 0.3 \\
\hline
\end{tabular}
\end{center}
\end{table}

\begin{figure}
   \centering
   \includegraphics[width=9cm]{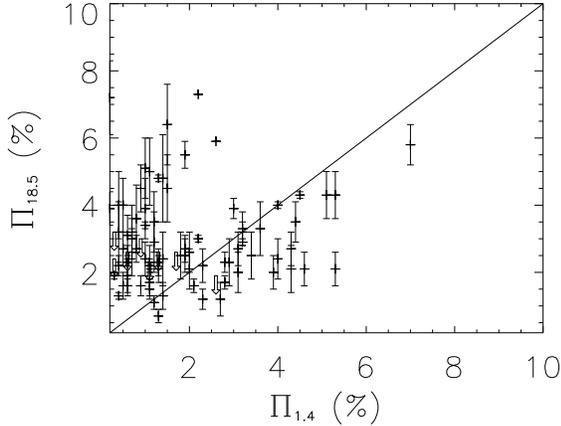}
\caption{$\Pi_{18.5}$ vs
$\Pi_{1.4}$ for the flat-spectrum sample. $\Pi_{18.5}$
upper limits are represented by downward arrows.}
\label{fig:FScorrPi}
\end{figure}

Unlike the steep-spectrum sources, we do not see an increase in
the fractional polarization
between 1.4 and 18.5 GHz.
As illustrated in Fig.~\ref{fig:FScorrPi}, sources that
are weakly polarized at 1.4 GHz show a stronger fractional polarization
at 18.5 GHz. This could be suggestive of differential Faraday
depolarization effects. Nevertheless, we do not see an increase of
the fractional polarization with redshift,
as might be expected in the case of Faraday depolarization inside
the source. It should be remembered that the emission
at different frequencies comes from different regions in the source,
and that these regions
may have intrinsically different polarization properties. 
On the other end, sources with $\Pi_{1.4} \gsim 2\%$
tend to have $\Pi_{1.4} \gsim \Pi_{18.5}$, which may be the result
of the decreased contribution, at high-frequency, of the more polarized,
steep-spectrum, diffuse component.      

\begin{figure}
   \centering
   \includegraphics[width=9cm]{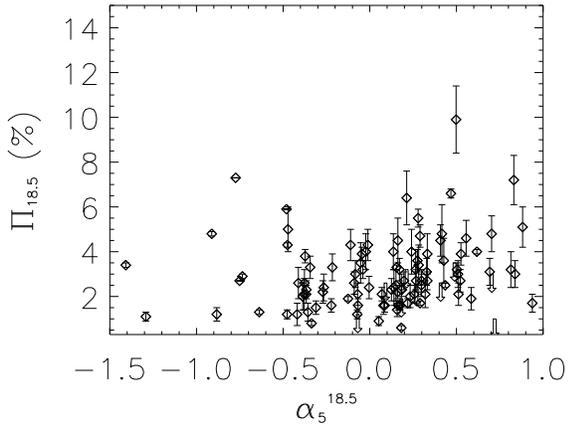}
\caption{$\Pi_{18.5}$ vs
$\alpha_{5}^{18.5}$ for the 108 flat-spectrum QSOs. $\Pi_{18.5}$
upper limits are indicated by downward arrows.} \label{fig:FScorr}
\end{figure}

To get further hints on mechanisms that determine the polarization
properties of sources, we used the generalized Kendall's rank
correlation coefficient Z to look for possible correlations
between $\Pi_{18.5}$, $\Pi_{1.4}$ and other properties of our
flat-spectrum radio sources, such as: the low-frequency spectral
index (between 1.4 GHz and 5 GHz), the high-frequency spectral
index (between 5 and 18.5 GHz), the monochromatic luminosity at 5
GHz. The values of the probability P that {\it no} correlation is
present are given for both the whole flat-spectrum sample and the
flat-spectrum QSOs only in Table~\ref{table:corrFS}. At variance
with what found for the steep-spectrum sample, there is no
evidence of a correlation between $\Pi_{18.5}$ and
$\alpha_{5}^{18.5}$ (see Fig.~\ref{fig:FScorr}).

\begin{table}
\begin{center}
\caption{Test for correlations between the fractional polarization at
18.5 and 1.4 GHz and other properties of the flat-spectrum
sources (whole sample and QSO sub-sample only). Z is the generalized
Kendall's correlation coefficient
and P is the probability that a correlation is not present.}
\label{table:corrFS}
\begin{tabular}{|l|rr|rr|}
\hline
 Correlation       & \multicolumn{2}{c|}{Z}  & \multicolumn{2}{c|}{P}   \\
\hline
\multicolumn{1}{|l|}{}  & \multicolumn{1}{c|}{All} & \multicolumn{1}{c|}{QSO}
                   & \multicolumn{1}{c|}{All} & \multicolumn{1}{c|}{QSO} \\
\hline
$\Pi_{18.5} - \alpha_{1.4}^{5}$   &  0.47   & 0.23   &  0.48 & 0.63 \\
$\Pi_{18.5} - \alpha_{5}^{18.5}$  &  1.83   & 2.29   &  0.07 & 0.02 \\
$\Pi_{18.5} - z   $               &  0.23   & 0.09   &  0.82 & 0.93  \\
$\Pi_{18.5} - L_{5\rm{GHz}}$      &  0.91   & 0.56   &  0.36 & 0.58 \\
$\Pi_{18.5} - \Pi_{1.4}$          &  0.95   & 0.24   &  0.34 & 0.81 \\
\hline
$\Pi_{1.4} - \alpha_{1.4}^{5}$   &  0.89   &  0.89  &  0.38 & 0.37 \\
$\Pi_{1.4} - \alpha_{5}^{18.5}$  &  1.90   &  1.47  &  0.05 & 0.14 \\
$\Pi_{1.4} -  z       $          &  1.85   &  1.53  &  0.06 & 0.13  \\
$\Pi_{1.4} - L_{5\rm{GHz}}$     &  1.75    &  1.58  &  0.08 & 0.11 \\
\hline
\end{tabular}
\end{center}
\end{table}

\section{Discussion and conclusions}

We have performed high-frequency observations of
a large, complete sample of bright extragalactic radio sources:
250 of the 258 sources in the southern sky ($\delta < 0^\circ$)
of the 5 GHz K\"uhr et al. (1981) sample were observed with the ATCA at
18.5 GHz. This sample is complete to
$S_{5{\rm GHz}} \ge 1\,$Jy. The polarized flux has
been measured to better than $5\sigma$ for 170 sources (114 flat-spectrum
and 56 steep-spectrum) and upper limits have been set for
additional 27 sources (12 flat-spectrum and 15 steep-spectrum).
This represents to 88\% of the flat-spectrum
sources and 62\% of the steep-spectrum sources in the southern K\"uhr sample.
53 sources were not considered in our analysis, as they appear to
be extended (i.e. our
point-source model fitting technique showed a significant
fraction of non-pointlike emission). Most of the extended sources
are steep-spectrum. Consequently our conclusions about this class of objects
may be biased
against extended sources.

The median fractional polarization at 18.5 GHz for steep and flat-spectrum
sources was found to be $\simeq 4.8\%$ and $\simeq 2.7\%$, respectively.
The distributions are both
skewed, with tails towards high values, up to a
maximum of $\simeq20\%$ and $\simeq 10\%$, respectively.
That steep-spectrum sources generally display higher fractional polarization
than flat-spectrum sources could be explained either in terms of
different intrinsic polarization properties of the two classes of objects,
or in terms of different internal depolarization effects (flat sources,
usually dominated by their core emission, could be more heavily depolarized).
An alternative to depolarization by internal Faraday rotation
is also offered by gradients in an external Faraday screen that
are not resolved by our observations.
However, these two depolarizing mechanisms 
cannot be distinguished given
polarization measurements at only two frequencies.
Moreover, spectral indices and polarization 
levels of low redshift
steep-spectrum radio sources with high polarization degrees 
($\Pi_{\rm 18.5} > 10\%$) could be overestimated due to resolution 
effect. Magnetic fields are more ordered on small angular scales, thus 
increasing polarization degrees on those scales.

We used the NVSS survey data to derive the fractional polarization at
1.4 GHz for the K\"uhr all-sky sample. We find very similar median
values for both the steep and the flat sources (1.1\% and 1.4\%,
respectively). This means a factor of $\sim$4 ($\sim$2) increase
from 1.4 to 18.5 GHz for the steep (flat) spectrum sources. It is
unclear whether such an increase can be attributed to a decreased
internal Faraday depolarization. In this case we would expect the
fractional polarization to be correlated with redshift. However such
a correlation is not found either for the steep or for the flat
spectrum sources. For steep-spectrum sources, any redshift
dependence could be counteracted by a steep increase of the
intrinsic Faraday rotation measures with redshift, reflecting an
increase of the environmental density (Pentericci et al., 2000).
For the flat-spectrum sample, on the other hand, it is more likely
that the increase in polarization is because the
radio emission at different frequencies comes from different core
components, that may have intrinsically different polarization
properties.

We do not see any significant correlation between $\Pi_{18.5}$ and
other source properties, such as the low-frequency spectral index and
the luminosity at 5 GHz. A possible weak (3$\sigma$) correlation
between $\Pi_{18.5}$ and the high-frequency spectral index,
$\alpha_5^{18.5}$, is apparent in the steep-spectrum sample only.

An important application of these results is in the estimate of the
contamination of CMB polarization maps by extragalactic radio
sources (Mesa et al. 2002). Based on the multifrequency radio
polarimetry by Klein et al. (2003), Mesa et al. (2002) estimated an
average increase by a factor of 3 of the fractional polarization of
steep-spectrum sources from 1.4 GHz to higher frequencies. Our data
suggest an increase by a larger factor ($\sim 4$), although caution is needed
because of the strong incompleteness of the steep-spectrum sample.
On the other hand, it is worth noticing that
steep-spectrum sources provide a minor contribution to
polarization fluctuations at frequencies ($\gsim 30\,$GHz) where
CMB measurements are carried out. For flat-spectrum sources,
Mesa et al. (2002) considered two possibilities: that the
fractional polarization remains, on average, constant from 1.4 GHz to
high frequencies (as found by Rudnick et al. 1985), or increases by a
factor of 3 (as found by Nartallo et al. 1998). Our results are
intermediate (an increase of a factor of $\sim 2$). This suggests
a contamination of CMB polarization maps which
is likely to be within the ranges considered by Mesa et
al. (2002).

\begin{acknowledgements}
      This research was supported in part by the Italian Space
      Agency (ASI) and by the Italian MIUR through a COFIN grant.
      RR gratefully acknowledges a financial contribution from the
      Italian National Research Council (CNR) as part of 
      an exchange program with CSIRO; he also warmly thanks the Paul Wild
      Observatory staff for
      their kind hospitality at Narrabri (NSW, Australia)
      where some of this work was completed.
      We are grateful to the referee for useful comments that helped
      improving the paper. 
\end{acknowledgements}


\begin{table*}
\begin{center}
\caption[]{Summary of observed and derived quantities of the steep
spectrum sample: (1) source name (B1950); (2) observation dates (1=March
20$-$21, 2= March 22, 3=March 27$-$28); (3) redshift; (4) source type;
 (5)--(6) 18.5 GHz total intensity and its
error in mJy; (7)--(8) 18.5 GHz polarized flux
and its error in mJy; (9)--(10)
18.5 GHz fractional polarization (\%) and its
error (\%); (11) spectral index between 1.4 and 5 GHz;
(12) spectral index between 5 and 18.5 GHz; (13) logarithm of the
5 GHz luminosity in W Hz$^{-1}$; (14) 1.4 GHz
NVSS fractional polarization (\%).}
\label{table:SStot}
\begin{tabular}{|c|c|r|c|r|r|r|r|r|r|r|r|r|r|}
\hline
  (1)  &       (2)       &    (3)  & (4) &  (5)           &    (6)      &      (7)         &  (8) &      (9)          &       (10)           &           (11)        &        (12)          &     (13)  &  (14)      \\
name   &    dates   & \multicolumn{1}{c|}{z}  &   type &
\multicolumn{1}{c|}{$I$}  & \multicolumn{1}{c|}{$\sigma_{I}$}   &
\multicolumn{1}{c|}{$F_{\rm p}$} &
\multicolumn{1}{c|}{$\sigma_{\rm F}$} &
\multicolumn{1}{c|}{$\Pi_{18.5}$} &
\multicolumn{1}{c|}{$\sigma_{\Pi}$} &
\multicolumn{1}{c|}{$\alpha_{1.4}^{5}$}  &
\multicolumn{1}{c|}{$\alpha_{5}^{18.5}$}
  &  \multicolumn{1}{c|}{$\log L_{\rm 5}$} &  \multicolumn{1}{c|}{$\Pi_{1.4}$}  \\
\hline
0003$-$003 &  2,3 &  1.03700 &    QSO &   525.0 &    14.0 & $<$43.5 &     8.7 &  $<$8.3  &   1.7 &    0.81 &    0.74 &   27.86 &     0.66  \\
0008$-$421 &    1 &    ...   &     EF &   181.6 &     0.6 &     9.4 &     1.2 &     5.2  &   0.7 &     ... &    1.46 &     ... &      ...  \\
0022$-$423 &    1 &    ...   &    QSO &   339.9 &     0.6 &    10.9 &     1.4 &     3.2  &   0.4 &     ... &    1.28 &     ... &      ...  \\
0023$-$263 &    3 &  0.32200 &    GAL &  1143.7 &     5.5 &    55.8 &     6.2 &     4.9  &   0.5 &    0.69 &    0.88 &   27.20 &     0.11  \\
0035$-$024 &  2,3 &  0.21970 &    GAL &   651.0 &    14.0 & $<$29.0 &     5.8 &  $<$4.5  &   0.9 &    0.65 &    1.08 &   26.73 &     0.20  \\
0039$-$445 &    1 &    ...   &    GAL &   218.6 &     0.7 &    15.7 &     2.0 &     7.2  &   0.9 &     ... &    1.30 &     ... &      ...  \\
0042$-$357 &    1 &    ...   &  NO ID &   269.9 &     0.7 &    20.6 &     2.1 &     7.6  &   0.8 &    0.73 &    1.00 &     ... &     1.31  \\
0045$-$255 &    2 &  0.00086 &    GAL &   638.4 &     2.9 & $<$16.6 &     3.3 &  $<$2.6  &   0.5 &    0.19 &    0.99 &   21.84 &  $<$0.10  \\
0105$-$163 &  2,3 &  0.40000 &    GAL &   114.0 &     2.0 &    14.2 &     2.6 &    12.5  &   2.3 &    0.99 &    1.78 &   26.95 &     0.45  \\
0114$-$211 &  2,3 &    ...   &     EF &   280.5 &     1.7 &    18.9 &     3.1 &     6.7  &   1.1 &    0.91 &    1.16 &     ... &     0.10  \\
0117$-$155 &  2,3 &  0.56500 &    GAL &   325.0 &     2.0 &    35.5 &     3.7 &    10.9  &   1.1 &    0.90 &    1.22 &   27.39 &     0.68  \\
0123$-$016 &  2,3 &  0.01770 &    GAL &   170.0 &     6.0 &    32.4 &     7.1 &    19.1  &   4.2 &     ... &    1.82 &     ... &      ...  \\
0131$-$367 &    1 &    ...   &    GAL &    54.9 &     0.6 &     7.8 &     1.3 &    14.1  &   2.4 &     ... &    2.83 &     ... &      ...  \\
0159$-$117 &  2,3 &  0.66900 &    QSO &   690.0 &     5.0 &    33.1 &     4.4 &     4.8  &   0.6 &    0.53 &    0.53 &   27.40 &     3.52  \\
0201$-$440 &    1 &    ...   &   QSO? &   256.0 &     1.4 &    37.3 &     2.4 &    14.6  &   0.9 &     ... &    1.07 &     ... &      ...  \\
0235$-$197 &    2 &  0.62000 &    GAL &   154.1 &     1.0 &    26.5 &     4.3 &    17.2  &   2.8 &    0.91 &    1.71 &   27.43 &     2.93  \\
0237$-$233 &    2 &  2.22400 &    QSO &   915.8 &     3.5 &    42.0 &     5.9 &     4.6  &   0.6 &    0.47 &    1.00 &   28.78 &     0.83  \\
0252$-$712 &    1 &    ...   &    GAL &   316.6 &     0.9 &    11.7 &     1.7 &     3.7  &   0.5 &     ... &    1.25 &     ... &      ...  \\
0407$-$658 &    1 &    ...   &   QSO? &   462.0 &     2.2 &    21.7 &     3.0 &     4.7  &   0.7 &     ... &    1.53 &     ... &      ...  \\
0409$-$752 &    1 &    ...   &    GAL &   984.0 &     4.8 &    73.3 &    11.7 &     7.5  &   1.2 &     ... &    1.15 &     ... &      ...  \\
0413$-$210 &    2 &  0.80700 &    QSO &   856.1 &     2.3 &    20.4 &     4.2 &     2.4  &   0.5 &    0.52 &    0.39 &   27.57 &     1.15  \\
0518$-$458 &    1 &    ...   &    GAL &   791.0 &     5.0 & $<$24.5 &     4.9 &  $<$3.1  &   0.6 &     ... &    2.27 &     ... &      ...  \\
0602$-$319 &    1 &  0.45200 &    QSO &   416.0 &     1.4 &    18.1 &     2.8 &     4.4  &   0.7 &    0.67 &    0.84 &   27.03 &     1.00  \\
0604$-$203 &    2 &  0.16400 &    GAL &   172.5 &     1.4 &    13.3 &     2.3 &     7.7  &   1.3 &    0.81 &    1.37 &   26.07 &     0.40  \\
0614$-$349 &    1 &  0.32900 &    GAL &   494.1 &     1.7 &    17.8 &     4.0 &     3.6  &   0.8 &    0.58 &    0.77 &   26.78 &  $<$0.08  \\
0806$-$103 &    2 &  0.10700 &    GAL &   388.0 &     3.0 &    18.4 &     4.7 &     4.7  &   1.2 &    0.40 &    1.10 &   25.87 &     4.10  \\
0834$-$196 &    2 &    ...   &   GAL? &   410.0 &     1.4 &    12.3 &     2.8 &     3.0  &   0.7 &    0.89 &    1.00 &     ... &     0.16  \\
0842$-$754 &    1 &    ...   &    QSO &   305.0 &     5.0 & $<$26.5 &     5.3 &  $<$8.7  &   1.7 &     ... &    1.17 &     ... &      ...  \\
0858$-$279 &    2 &  2.15200 &    QSO &  1477.0 &     1.6 &   105.5 &     6.7 &     7.1  &   0.5 &    0.02 & $-$0.03 &   28.15 &  $<$0.15  \\
0915$-$118 &    2 &  0.05470 &    GAL &  1851.4 &     3.3 &   222.4 &    18.8 &    12.0  &   1.0 &    0.84 &    1.54 &   26.23 &     0.04  \\
0941$-$080 &    2 &    ...   &    GAL &   371.5 &     2.1 & $<$12.7 &     2.5 &  $<$3.4  &   0.7 &    0.72 &    0.82 &     ... &     0.09  \\
1015$-$314 &    1 &    ...   &    GAL &   451.3 &     1.6 &    37.8 &     3.4 &     8.4  &   0.8 &    0.79 &    0.86 &     ... &     0.08  \\
1017$-$426 &    1 &    ...   &    QSO &   249.9 &     0.7 &    21.2 &     1.8 &     8.5  &   0.7 &     ... &    1.24 &     ... &      ...  \\
1143$-$483 &    1 &    ...   &    GAL &   419.7 &     1.9 & $<$15.8 &     3.2 &  $<$3.8  &   0.8 &     ... &    0.82 &     ... &      ...  \\
1151$-$348 &    1 &  0.25800 &    QSO &   996.9 &     1.8 &    27.0 &     5.0 &     2.7  &   0.5 &    0.61 &    0.79 &   26.89 &  $<$0.04  \\
1221$-$423 &    1 &    ...   &    GAL &   359.5 &     1.2 &    18.9 &     1.9 &     5.3  &   0.5 &     ... &    0.80 &     ... &      ...  \\
1229$-$021 &  2,3 &  1.04500 &    QSO &   679.7 &     2.1 &    26.2 &     6.9 &     3.8  &   1.0 &    0.33 &    0.34 &   27.60 &     0.58  \\
1239$-$044 &  2,3 &  0.48000 &    GAL &   186.0 &     3.0 & $<$17.5 &     3.5 &  $<$9.4  &   1.9 &    1.01 &    1.29 &   27.05 &     0.45  \\
1245$-$197 &    2 &  1.27500 &    QSO &   767.7 &     1.6 &    16.9 &     4.3 &     2.2  &   0.6 &    0.56 &    0.90 &   28.22 &     0.33  \\
1246$-$410 &    1 &    ...   &    GAL &   209.1 &     0.9 &    23.4 &     2.7 &    11.2  &   1.3 &     ... &    1.44 &     ... &      ...  \\
1306$-$095 &  2,3 &  0.46400 &   GAL? &   841.0 &     3.0 &    54.5 &    13.9 &     6.5  &   1.7 &    0.61 &    0.63 &   27.24 &     0.28  \\
1308$-$220 &    2 &    ...   &     EF &   215.8 &     1.3 &    21.6 &     3.4 &    10.0  &   1.6 &    0.99 &    1.28 &     ... &     0.34  \\
1318$-$434 &    1 &    ...   &    GAL &   523.4 &     1.5 & $<$12.9 &     2.6 &  $<$2.5  &   0.5 &     ... &    1.03 &     ... &      ...  \\
1320$-$446 &    1 &    ...   &    QSO &   283.1 &     0.6 &     8.9 &     1.3 &     3.2  &   0.5 &     ... &    1.01 &     ... &      ...  \\
1322$-$428 &    1 &    ...   &    GAL &  6609.9 &     6.9 &    93.0 &     3.5 &     1.4  &   0.1 &     ... & $-$0.60 &     ... &      ...  \\
1331$-$098 &  2,3 &  0.08100 &    GAL &    70.0 &     4.0 & $<$22.5 &     4.5 & $<$32.1  &   6.5 &     ... &    2.25 &     ... &      ...  \\
1333$-$337 &    1 &  0.01290 &    GAL &   269.7 &     1.7 &    16.8 &     2.3 &     6.2  &   0.9 &     ... &    2.47 &     ... &      ...  \\
1335$-$061 &  2,3 &  0.62500 &    QSO &   146.0 &     3.0 & $<$13.5 &     2.7 &  $<$9.2  &   1.9 &    0.83 &    1.48 &   27.27 &     2.24  \\
1355$-$416 &    1 &    ...   &    QSO &   194.5 &     1.7 &    11.3 &     1.9 &     5.8  &   1.0 &     ... &    1.53 &     ... &      ...  \\
1524$-$136 &  2,3 &  1.68700 &    QSO &   461.4 &     3.7 &    20.5 &     7.6 &     4.4  &   1.6 &    0.66 &    0.74 &   28.19 &     0.22  \\
1637$-$771 &    1 &    ...   &    GAL &   430.8 &     1.7 & $<$18.0 &     3.6 &  $<$4.2  &   0.8 &     ... &    1.44 &     ... &      ...  \\
1733$-$565 &    1 &    ...   &    GAL &   342.5 &     1.7 &    14.3 &     2.4 &     4.2  &   0.7 &     ... &    1.76 &     ... &      ...  \\
1740$-$517 &    1 &    ...   &   GAL? &  1361.8 &     1.6 &    16.5 &     3.1 &     1.2  &   0.2 &     ... &    0.61 &     ... &      ...  \\
1814$-$637 &    1 &    ...   &    GAL &  1584.3 &     2.2 & $<$13.0 &     2.6 &  $<$0.8  &   0.2 &     ... &    0.79 &     ... &      ...  \\
1829$-$718 &    1 &    ...   &  NO ID &   258.3 &     0.5 &    10.0 &     1.2 &     3.9  &   0.4 &     ... &    1.05 &     ... &      ...  \\
1934$-$638 &1,2,3 &    ...   &    GAL &  1166.3 &     0.3 &     8.6 &     0.5 &     0.7  &   0.0 &     ... &    1.28 &     ... &      ...  \\
1938$-$155 &    3 &  0.45200 &    GAL &   495.1 &     7.2 &    39.9 &     9.0 &     8.1  &   1.8 &    0.86 &    1.15 &   27.32 &     3.32  \\
2008$-$068 &    3 &    ...   &    GAL &   399.7 &     6.2 &    28.5 &     7.0 &     7.1  &   1.8 &    0.50 &    0.94 &     ... &     0.28  \\
2032$-$350 &    1 &    ...   &   GAL? &   468.0 &     2.0 &    17.9 &     6.0 &     3.8  &   1.3 &    0.84 &    1.08 &     ... &     5.75  \\
2044$-$027 &    3 &  0.94200 &    QSO &   382.0 &     6.0 & $<$27.0 &     5.4 &  $<$7.1  &   1.4 &    0.63 &    0.75 &   27.58 &     0.35  \\
2053$-$201 &    3 &  0.15600 &    GAL &   109.0 &     7.0 &    38.4 &     8.1 &    35.2  &   7.5 &    0.77 &    1.70 &   26.01 &  $<$0.14  \\
2135$-$209 &    3 &  0.63500 &    GAL &   479.9 &     7.1 &    39.4 &    10.0 &     8.2  &   2.1 &    0.69 &    0.89 &   27.43 &  $<$0.06  \\
\hline
\end{tabular}
\end{center}
\end{table*}

\begin{table*}
\noindent Table 5 ({\it continued})
\begin{center}
\begin{tabular}{|c|c|r|c|r|r|r|r|r|r|r|r|r|r|}
\hline
  (1)  &       (2)       &    (3)  & (4) &  (5)           &    (6)      &      (7)         &  (8) &      (9)          &       (10)           &           (11)        &        (12)          &     (13)  &  (14)      \\
name   &    dates   & \multicolumn{1}{c|}{z}  &   type &
\multicolumn{1}{c|}{$I$}  & \multicolumn{1}{c|}{$\sigma_{I}$}   &
\multicolumn{1}{c|}{$F_{\rm p}$} &
\multicolumn{1}{c|}{$\sigma_{\rm F}$} &
\multicolumn{1}{c|}{$\Pi_{18.5}$} &
\multicolumn{1}{c|}{$\sigma_{\Pi}$} &
\multicolumn{1}{c|}{$\alpha_{1.4}^{5}$}  &
\multicolumn{1}{c|}{$\alpha_{5}^{18.5}$}
  &  \multicolumn{1}{c|}{$\log L_{\rm 5}$} &  \multicolumn{1}{c|}{$\Pi_{1.4}$}  \\
\hline
2140$-$816 &   1 & ... & NO ID & 149.8  & 1.4  &    12.3  & 1.9  &     8.2 &  1.3 &  ...  & 1.451  &  ... &     ...  \\
2149$-$287 &   3 & ... & NO ID & 403.6  & 8.1  &    43.3  & 9.4  &    10.7 &  2.3 & 0.585 & 0.929  &  ... &    2.67  \\
2150$-$520 &   1 & ... &  QSO? & 268.9  & 0.6  &    11.2  & 1.8  &     4.2 &  0.7 &  ...  & 1.150  &  ... &     ...  \\
2226$-$411 &   1 & ... &   QSO & 295.2  & 0.8  &    25.4  & 1.5  &     8.6 &  0.5 &  ...  & 0.991  &  ... &     ...  \\
2250$-$412 &   1 & ... &   GAL & 281.3  & 0.7  &    13.7  & 2.0  &     4.9 &  0.7 &  ...  & 1.187  &  ... &     ...  \\
2252$-$530 &   1 & ... &  GAL? & 279.2  & 0.6  &    17.5  & 1.3  &     6.3 &  0.5 &  ...  & 0.983  &  ... &     ...  \\
2317$-$277 & 2,3 & ... &   GAL &  67.0  & 2.0  & $<$32.0  & 6.4  & $<$47.8 &  9.6 & 0.398 & 2.267  &  ... &   16.20  \\
2323$-$407 &   1 & ... &   GAL & 221.5  & 0.8  &    40.5  & 1.5  &    18.3 &  0.7 &  ...  & 1.218  &  ... &     ...  \\
2331$-$417 &   1 & ... &   GAL & 290.0  & 1.4  &    12.8  & 2.3  &     4.4 &  0.8 &  ...  & 1.301  &  ... &     ...  \\
\hline
\end{tabular}
\end{center}
\end{table*}



\begin{table*}
\begin{center}
\caption{Summary of observed and derived quantities for
flat-spectrum sources. Same columns as in
Table~\ref{table:SStot}.} \label{table:FStot}
\begin{tabular}{|c|c|r|c|r|r|r|r|r|r|r|r|r|r|}
\hline
  (1)  &       (2)       &    (3)  & (4) &  (5)           &    (6)      &      (7)         &  (8) &      (9)          &       (10)           &           (11)        &        (12)          &     (13)  &  (14)      \\
name   &    dates   & \multicolumn{1}{c|}{z}  &   type &
\multicolumn{1}{c|}{$I$}  & \multicolumn{1}{c|}{$\sigma_{I}$}   &
\multicolumn{1}{c|}{$F_{\rm p}$} &
\multicolumn{1}{c|}{$\sigma_{\rm F}$} &
\multicolumn{1}{c|}{$\Pi_{18.5}$} &
\multicolumn{1}{c|}{$\sigma_{\Pi}$} &
\multicolumn{1}{c|}{$\alpha_{1.4}^{5}$}  &  \multicolumn{1}{c|}{$\alpha_{5}^{18.5}$}
  &  \multicolumn{1}{c|}{$\log L_{\rm 5}$} &  \multicolumn{1}{c|}{$\Pi_{1.4}$}  \\
\hline
0003$-$066 &  2,3   &     0.34700 &        QSO & 2561.3 &  11.9 &   29.8  &  13.4 &  1.2   &  0.5 &    0.25 & $-$0.41 & 26.82 &  2.7 \\
0047$-$579 &  1     &     1.79700 &        QSO & 1643.0 &   2.8 &   77.5  &   7.8 &  4.7   &  0.5 &    ...  &    0.29 &  ...  &  ... \\
0048$-$097 &  2,3   &       ...   &        QSO & 1397.1 &   4.9 &   37.2  &   7.7 &  2.7   &  0.5 & $-$0.69 &    0.26 &  ...  &  3.1 \\
0056$-$572 &  1     &       ...   &        QSO &  580.1 &   3.1 &$<$20.1  &   4.0 & $<$3.5 &  0.7 &    ...  &    0.48 &  ...  &  ... \\
0112$-$017 &  2,3   &     1.38100 &        QSO &  777.6 &   5.4 &   30.5  &   6.9 &  3.9   &  0.9 & $-$0.08 &    0.33 & 27.72 &  1.0 \\
0113$-$118 &  2,3   &     0.67200 &        QSO & 1430.1 &   2.9 &   28.4  &   7.6 &  2.0   &  0.5 & $-$0.06 &    0.23 & 27.41 &  3.9 \\
0122$-$003 &  2,3   &     1.07000 &        GAL & 1721.5 &   3.4 &   60.0  &  10.7 &  3.5   &  0.6 &    0.17 & $-$0.25 & 27.63 &  4.4 \\
0130$-$171 &  2,3   &     1.02200 &        QSO & 1614.4 &   8.4 &   36.7  &   6.8 &  2.3   &  0.4 & $-$0.14 & $-$0.36 & 27.41 &  1.3 \\
0131$-$522 &  1     &       ...   &        QSO &  794.5 &   1.8 &   24.7  &   4.5 &  3.1   &  0.6 &    ...  &    0.32 &  ...  &  ... \\
0138$-$097 &  2,3   &     0.50100 &        QSO &  922.8 &   3.0 &   59.4  &  11.3 &  6.4   &  1.2 & $-$0.48 &    0.21 &  ...  &  1.5 \\
0202$-$172 &  2     &     1.74000 &        QSO & 1087.3 &   4.8 &$<$20.3  &   4.1 & $<$1.9 &  0.4 & $-$0.09 &    0.18 & 27.93 &  2.6 \\
0208$-$512 &  1     &     1.00300 &     BL/QSO & 2913.7 &   7.1 &   99.2  &   9.9 &  3.4   &  0.3 &    ...  &    0.09 &  ...  &  ... \\
0238$-$084 &  2,3   &     0.00500 &        QSO & 1012.5 &   8.2 &$<$24.4  &   4.9 & $<$2.4 &  0.5 & $-$0.35 &    0.26 &  ...  &  0.3 \\
0302$-$623 &  1     &       ...   &        QSO & 1870.5 &   2.4 &   61.8  &  10.9 &  3.3   &  0.6 &    ...  & $-$0.21 &  ...  &  ... \\
0308$-$611 &  1     &       ...   &        QSO & 1215.5 &   1.0 &   18.9  &   4.1 &  1.6   &  0.3 &    ...  &    0.08 &  ...  &  ... \\
0332$-$403 &  1     &     1.44500 &        QSO & 1786.4 &   1.9 &   63.9  &   9.1 &  3.6   &  0.5 &    ...  &    0.27 &  ...  &  ... \\
0336$-$019 &  2,3   &     0.85200 &        QSO & 4485.9 &   7.8 &  147.1  &  21.7 &  3.3   &  0.5 & $-$0.13 & $-$0.34 & 27.74 &  3.2 \\
0400$-$319 &  1     &       ...   &        GAL &  885.6 &   3.2 &   22.0  &   5.8 &  2.5   &  0.7 & $-$0.34 &    0.13 &  ...  &  1.8 \\
0402$-$362 &  1     &     1.41700 &        QSO & 3180.9 &   3.0 &   42.2  &   4.1 &  1.3   &  0.1 & $-$0.14 & $-$0.63 & 27.78 &  0.4 \\
0403$-$132 &  2     &     0.57100 &        QSO & 2292.2 &   3.0 &   36.6  &   5.6 &  1.6   &  0.2 &    0.30 &    0.17 & 27.53 &  2.1 \\
0405$-$385 &  1     &     1.28500 &        QSO & 1636.5 &   2.9 &   24.9  &   5.7 &  1.5   &  0.3 & $-$0.18 & $-$0.31 & 27.59 &  1.1 \\
0405$-$123 &  2     &     0.57400 &     BL/QSO & 1430.6 &   3.3 &   23.8  &   5.8 &  1.7   &  0.4 &    0.31 &    0.24 & 27.37 &  1.1 \\
0414$-$189 &  2     &     1.53600 &        QSO &  913.0 &   2.8 &   25.1  &   4.4 &  2.7   &  0.5 & $-$0.06 &    0.29 & 27.85 &  0.5 \\
0420$-$014 &  2     &     0.91500 &        QSO & 7912.8 &  13.0 &   85.5  &  12.5 &  1.1   &  0.2 &    0.49 & $-$1.29 & 27.67 &  1.2 \\
0426$-$380 &  1     &     1.03000 &     BL/QSO & 1056.3 &   2.2 &   61.5  &   6.2 &  5.8   &  0.6 & $-$0.36 &    0.07 &  ...  &  7.0 \\
0434$-$188 &  2     &     2.70200 &        QSO &  423.9 &   2.1 &   13.7  &   3.6 &  3.2   &  0.8 & $-$0.42 &    0.81 & 27.98 &  0.7 \\
0437$-$454 &  1     &       ...   &        QSO & 1127.2 &   1.4 &   36.5  &   5.1 &  3.2   &  0.5 &    ...  &    0.17 &  ...  &  ... \\
0438$-$436 &  1     &     2.85200 &        QSO & 3926.9 &   3.0 &   97.3  &   2.4 &  2.5   &  0.1 &    ...  &    0.43 &  ...  &  ... \\
0440$-$003 &  2     &     0.84400 &        QSO & 1214.8 &   3.4 &   22.6  &   6.4 &  1.9   &  0.5 & $-$0.30 &    0.58 & 27.65 &  0.6 \\
0454$-$810 &  1     &     0.44400 &        QSO & 1574.3 &   3.5 &   40.7  &   7.3 &  2.6   &  0.5 &    ...  & $-$0.09 &  ...  &  ... \\
0451$-$282 &  2     &     2.55900 &        QSO & 1746.6 &   1.6 &   46.9  &   7.8 &  2.7   &  0.4 &    0.09 &    0.19 & 28.50 &  0.8 \\
0454$-$463 &  1     &     0.85800 &        QSO & 3599.6 &   2.3 &   27.8  &   3.8 &  0.8   &  0.1 &    ...  & $-$0.33 &  ...  &  ... \\
0454$-$234 &  2     &     1.00300 &        QSO & 5389.8 &   4.9 &  157.5  &   4.3 &  2.9   &  0.1 & $-$0.13 & $-$0.73 & 27.71 &  3.2 \\
0458$-$020 &  2     &     2.28600 &        QSO & 1568.7 &   3.5 &   24.5  &   6.8 &  1.6   &  0.4 &    0.20 &    0.07 & 28.37 &  0.5 \\
0506$-$612 &  1     &     1.09300 &     BL/QSO & 2436.4 &   2.0 &   50.2  &   9.5 &  2.1   &  0.4 &    ...  & $-$0.26 &  ...  &  ... \\
0511$-$220 &  2     &     1.29600 &        QSO &  768.5 &   2.2 &   34.2  &   5.7 &  4.5   &  0.7 & $-$0.55 &    0.40 & 27.54 &  0.9 \\
0514$-$459 &  1     &     0.19400 &        QSO &  886.8 &   2.1 &   35.9  &   7.1 &  4.0   &  0.8 &    ...  &    0.13 &  ...  &  ... \\
0524$-$460 &  1     &     1.47900 &        QSO &  523.9 &   2.6 &   15.9  &   3.3 &  3.0   &  0.6 &    ...  &    0.50 &  ...  &  ... \\
0528$-$250 &  2     &     2.76500 &        QSO &  469.3 &   2.7 &   14.4  &   2.7 &  3.1   &  0.6 &    0.00 &    0.69 & 28.21 &  0.6 \\
0537$-$441 &  1     &     0.89600 &        QSO &10682.9 &   3.9 &  290.8  &   2.7 &  2.7   &  0.0 &    ...  & $-$0.75 &  ...  &  ... \\
0537$-$286 &  2     &     3.11900 &        QSO &  865.7 &   3.0 &   20.1  &   3.9 &  2.3   &  0.5 & $-$0.13 &    0.12 & 28.16 &  0.6 \\
0539$-$057 &  2     &     0.83900 &        QSO &  782.7 &   2.8 &   21.0  &   4.1 &  2.7   &  0.5 & $-$0.44 &    0.52 & 27.38 &  4.3 \\
0605$-$085 &  2     &     0.87200 &     BL/QSO & 1958.8 &   3.2 &   58.6  &   2.8 &  3.0   &  0.1 & $-$0.47 &    0.44 & 27.75 &  2.2 \\
0606$-$223 &  2     &     1.92600 &        QSO &  949.1 &   1.8 &   23.6  &   4.6 &  2.5   &  0.5 & $-$0.56 &    0.29 & 27.79 &  1.9 \\
0607$-$157 &  2     &     0.32400 &        QSO & 5999.5 &   7.5 &  287.2  &   5.1 &  4.8   &  0.1 &    0.32 & $-$0.91 & 26.86 &  1.3 \\
0637$-$752 &  1     &     0.65400 &        QSO & 4777.8 &   4.4 &  156.8  &   3.0 &  3.3   &  0.1 &    ...  &    0.15 &  ...  &  ... \\
0642$-$349 &  1     &     2.16500 &        QSO &  321.5 &   2.1 &   16.5  &   3.0 &  5.1   &  0.9 & $-$0.27 &    0.88 & 27.86 &  1.0 \\
0743$-$006 &  2     &     0.99400 &        QSO & 1615.1 &   3.7 &   35.1  &   5.1 &  2.2   &  0.3 & $-$0.86 &    0.16 & 27.47 &  0.4 \\
0805$-$077 &  2     &     1.83700 &        QSO & 1659.5 &   2.0 &   21.0  &   6.2 &  1.3   &  0.4 &    0.33 & $-$0.35 & 28.04 &  1.4 \\
0834$-$201 &  2     &     2.75200 &        QSO & 4383.0 &   5.8 &   81.7  &   3.9 &  1.9   &  0.1 & $-$0.49 & $-$0.12 & 28.43 &  0.3 \\
0859$-$140 &  2     &     1.33900 &        QSO & 1190.3 &   1.5 &   38.4  &   3.5 &  3.2   &  0.3 &    0.18 &    0.50 & 28.08 &  3.2 \\
0919$-$260 &  2     &     2.30000 &        QSO & 1442.5 &   2.1 &   23.0  &   4.7 &  1.6   &  0.3 & $-$0.01 & $-$0.06 & 28.14 &  0.9 \\
1032$-$199 &  2     &     2.19800 &        QSO & 1260.3 &   2.4 &   26.7  &   6.5 &  2.1   &  0.5 & $-$0.18 & $-$0.07 & 27.97 &  5.3 \\
1045$-$188 &  2     &     0.59500 &        QSO & 1625.2 &   2.1 &   36.4  &   7.7 &  2.2   &  0.5 &    0.01 & $-$0.27 & 27.10 &  2.3 \\
1057$-$797 &  1     &       ...   &        QSO & 2638.0 &   1.9 &  100.1  &   8.4 &  3.8   &  0.3 &    ...  & $-$0.37 &  ...  &  ... \\
1104$-$445 &  1     &     1.59800 &        QSO & 2273.0 &   2.4 &   27.7  &   3.5 &  1.2   &  0.2 &    ...  & $-$0.07 &  ...  &  ... \\
1116$-$462 &  1     &     0.71300 &     BL/QSO &  963.8 &   2.3 &   39.8  &   1.3 &  4.1   &  0.1 &    ...  &    0.25 &  ...  &  ... \\
1127$-$145 &  2     &     1.18700 &        QSO & 2926.2 &   2.2 &  116.3  &   2.7 &  4.0   &  0.1 & $-$0.12 &    0.61 & 28.34 &  4.0 \\
1143$-$245 &  2     &     1.95000 &        QSO &  691.3 &   3.3 &$<$17.7  &   3.5 & $<$2.6 &  0.5 &    0.15 &    0.40 & 28.06 &  1.7 \\
1144$-$379 &  1     &     1.04800 &        QSO & 4431.6 &   5.8 &  323.7  &   2.2 &  7.3   &  0.0 &    0.08 & $-$0.77 &  ...  &  2.2 \\
1145$-$071 &  2     &   1.34200  &       QSO &  809.9 &   1.7 &   22.2 &    3.6 &    2.7 &  0.4 & $-$0.41 &    0.33 & 27.60 &    1.9  \\
1148$-$001 &  2     &   1.98200  &       QSO &  972.2 &   1.0 &   20.6 &    4.6 &    2.1 &  0.5 &    0.29 &    0.51 & 28.35 &    4.6  \\
1202$-$262 &  2     &   0.78900  &     NO ID &  509.3 &   1.5 &$<$11.5 &    2.3 & $<$2.3 &  0.5 &    0.36 &    0.54 & 27.37 &    1.1  \\
1213$-$172 &  2     &     ...    &       GAL & 1897.1 &   2.5 &   41.3 &    3.9 &    2.2 &  0.2 &    0.09 & $-$0.19 &  ...  &    1.1  \\
1237$-$101 &  2,3   &   0.75300  &       QSO & 1032.6 &   5.4 &   23.6 &    7.6 &    2.3 &  0.7 &    0.11 &    0.18 & 27.37 &    2.9  \\
1243$-$072 &  2,3   &   1.28600  &       QSO &  712.1 &   1.0 &   24.5 &    6.7 &    3.4 &  0.9 &    ...  &    0.28 &  ...  &    ...  \\
\hline
\end{tabular}
\end{center}
\end{table*}

\begin{table*}
\noindent Table 6 ({\it continued})
\begin{center}
\begin{tabular}{|c|c|r|c|r|r|r|r|r|r|r|r|r|r|}
\hline
  (1)  &       (2)       &    (3)  & (4) &  (5)           &    (6)      &      (7)         &  (8) &      (9)          &       (10)           &           (11)        &        (12)          &     (13)  &  (14)      \\
name   &    dates   & \multicolumn{1}{c|}{z}  &   type &
\multicolumn{1}{c|}{$I$}  & \multicolumn{1}{c|}{$\sigma_{I}$}   &
\multicolumn{1}{c|}{$F_{\rm p}$} &
\multicolumn{1}{c|}{$\sigma_{\rm F}$} &
\multicolumn{1}{c|}{$\Pi_{18.5}$} &
\multicolumn{1}{c|}{$\sigma_{\Pi}$} &
\multicolumn{1}{c|}{$\alpha_{1.4}^{5}$}  &
\multicolumn{1}{c|}{$\alpha_{5}^{18.5}$}
  &  \multicolumn{1}{c|}{$\log L_{\rm 5}$} &  \multicolumn{1}{c|}{$\Pi_{1.4}$}  \\
\hline
1244$-$255 &  2     &   0.63300  &       GAL & 1848.1 &   3.1 &   60.6 &   14.4 &    3.3 &  0.8 & $-$0.12 & $-$0.23 & 27.20 &    3.6  \\
1253$-$055 &  1,2,3 &   0.53600  &       QSO &28043.5 &  12.3 & 1648.5 &    1.7 &    5.9 &  0.0 & $-$0.33 & $-$0.48 & 28.07 &    2.6  \\
1255$-$316 &  1     &   1.92400  &       QSO & 2000.6 &   3.5 &   86.9 &   13.7 &    4.3 &  0.7 & $-$0.32 & $-$0.11 & 27.99 &    5.3  \\
1302$-$102 &  2,3   &   0.28600  &       QSO &  827.0 &   2.0 &   26.4 &    7.4 &    3.2 &  0.9 & $-$0.39 &    0.26 & 26.48 &    0.4  \\
1313$-$333 &  1     &   1.21000  &    BL/QSO & 1457.6 &   3.0 &   33.2 &    8.0 &    2.3 &  0.5 & $-$0.04 & $-$0.05 & 27.69 &    1.1  \\
1334$-$127 &  2,3   &   0.53900  &       QSO & 7107.4 &   6.8 &   88.4 &   23.0 &    1.2 &  0.3 &    0.14 & $-$0.88 & 27.34 &    2.3  \\
1352$-$104 &  2,3   &   0.33200  &       QSO & 1659.4 &   2.0 &   35.1 &   10.9 &    2.1 &  0.7 & $-$0.21 & $-$0.37 & 26.56 &    4.3  \\
1354$-$152 &  2     &   1.89000  &       QSO &  999.8 &   1.3 &   21.4 &    6.0 &    2.1 &  0.6 & $-$0.63 &    0.32 & 27.78 &    2.0  \\
1406$-$076 &  2,3   &   1.49400  &       QSO & 1792.3 &   4.9 &   35.2 &   11.3 &    2.0 &  0.6 & $-$0.35 & $-$0.38 & 27.62 &    3.1  \\
1424$-$418 &  1     &   1.52200  &       QSO & 2923.3 &   2.1 &   25.6 &    4.5 &    0.9 &  0.2 &    ...  &    0.05 &  ...  &    ...  \\
1451$-$375 &  1     &   0.31400  &       QSO & 2004.8 &   3.1 &   78.1 &    5.8 &    3.9 &  0.3 & $-$0.54 & $-$0.04 & 26.75 &    3.0  \\
1504$-$166 &  3     &   0.87600  &       QSO & 1498.6 &  10.0 &$<$47.3 &    9.5 & $<$3.2 &  0.6 &    0.24 &    0.21 & 27.70 &    0.3  \\
1508$-$055 &  2,3   &   1.19100  &    BL/QSO & 1272.2 &   6.2 &   37.3 &   11.9 &    2.9 &  0.9 &    0.30 &    0.49 & 28.05 &    1.2  \\
1510$-$089 &  2,3   &   0.36100  &       QSO & 2527.6 &   5.7 &   63.1 &   18.9 &    2.5 &  0.7 & $-$0.10 &    0.15 & 27.12 &    3.4  \\
1514$-$241 &  3     &   0.04860  &       QSO & 2030.1 &  11.4 &   86.9 &   13.4 &    4.3 &  0.7 &    0.01 & $-$0.01 &  ...  &    5.1  \\
1519$-$273 &  3     &     ...    &       QSO & 1632.3 &   8.8 &   90.4 &    6.8 &    5.5 &  0.4 & $-$0.59 &    0.27 &  ...  &    1.9  \\
1541$-$828 &  1     &     ...    &       GAL &  589.7 &   1.8 &   15.0 &    3.1 &    2.5 &  0.5 &    ...  &    0.69 &  ...  &    ...  \\
1549$-$790 &  1     &   0.14900  &       QSO & 1429.2 &   2.1 &$<$14.8 &    3.0 & $<$1.0 &  0.2 &    ...  &    0.72 &  ...  &    ...  \\
1555+001   &  2,3   &   1.77000  &       QSO &  891.5 &   3.8 &$<$26.3 &    5.3 & $<$3.0 &  0.6 & $-$0.55 &    0.70 & 27.95 &    0.9  \\
1610$-$771 &  1     &   1.71000  &       QSO & 2674.3 &   3.0 &   46.1 &    2.0 &    1.7 &  0.1 &    ...  &    0.29 &  ...  &    ...  \\
1619$-$680 &  1     &   1.35400  &       GAL &  724.7 &   1.4 &$<$14.7 &    2.9 & $<$2.0 &  0.4 &    ...  &    0.72 &  ...  &    ...  \\
1622$-$253 &  3     &     ...    &       QSO & 2144.1 &  19.4 &   86.7 &   16.8 &    4.0 &  0.8 &    0.15 & $-$0.02 &  ...  &    0.5  \\
1718$-$649 &  1     &   0.01450  &       QSO & 3003.2 &   2.7 &   18.7 &    4.4 &    0.6 &  0.1 &    ...  &    0.18 &  ...  &      ...\\
1741$-$038 &  3     &   1.05700  &       QSO & 4921.9 &  21.8 &   77.2 &   13.6 &    1.6 &  0.3 & $-$0.75 & $-$0.22 & 27.81 &    0.6  \\
1815$-$553 &  1     &     ...    &       QSO & 1097.2 &   2.2 &   15.4 &    3.4 &    1.4 &  0.3 &    ...  &    0.15 &  ...  &    ...  \\
1831$-$711 &  1     &   1.35600  &       QSO & 2221.1 &   2.1 &   27.1 &    4.3 &    1.2 &  0.2 &    ...  & $-$0.47 &  ...  &    ...  \\
1903$-$802 &  1     &   0.50000  &       QSO &  614.9 &   2.0 &   18.5 &    3.5 &    3.0 &  0.6 &    ...  &    0.83 &  ...  &    ...  \\
1933$-$400 &  1     &   0.96600  &       QSO & 1162.7 &   1.5 &   26.9 &    3.3 &    2.3 &  0.3 & $-$0.30 &    0.18 & 27.49 &    2.8  \\
1936$-$155 &  3     &   1.65700  &       QSO &  966.6 &   8.1 &   34.7 &    9.4 &    3.6 &  1.0 & $-$0.80 &    0.42 & 27.69 &    0.8  \\
1936$-$623 &  1     &     ...    &       QSO &  531.4 &   1.4 &   24.4 &    4.0 &    4.6 &  0.8 &    ...  &    0.55 &  ...  &    ...  \\
1954$-$388 &  1     &   0.62600  &       GAL & 3810.4 &   1.8 &   26.5 &    6.0 &    0.7 &  0.2 & $-$0.25 & $-$0.47 & 27.34 &    1.3  \\
1958$-$179 &  3     &   0.65000  &       QSO & 1337.7 &   6.0 &   39.7 &    8.7 &    3.0 &  0.6 & $-$0.61 & $-$0.08 & 27.06 &    0.7  \\
2000$-$330 &  1     &   3.77700  &        EF &  514.7 &   1.9 &$<$13.3 &    2.7 & $<$2.6 &  0.5 & $-$0.74 &    0.61 & 27.91 & $<$0.6  \\
2005$-$489 &  1     &   0.07100  &       QSO & 1345.5 &   1.3 &$<$15.9 &    3.2 & $<$1.2 &  0.2 &    ...  & $-$0.06 &  ...  &    ...  \\
2008$-$159 &  3     &   1.18000  &       QSO & 2245.9 &   7.5 &   46.7 &    9.3 &    2.1 &  0.4 & $-$0.73 & $-$0.36 & 27.45 &    1.3  \\
2052$-$474 &  1     &   1.49100  &       QSO &  739.1 &   1.5 &   12.3 &    2.8 &    1.7 &  0.4 &    ...  &    0.93 &  ...  &    ...  \\
2106$-$413 &  1     &   1.05470  &       QSO & 1895.0 &   2.8 &   29.7 &    4.5 &    1.6 &  0.2 &    ...  &    0.16 &  ...  &    ...  \\
2126$-$158 &  3     &   3.26600  &       QSO &  933.5 &  10.9 &   36.9 &    9.5 &    4.0 &  1.0 & $-$0.60 &    0.24 & 27.98 & $<$0.4  \\
2128$-$123 &  3     &   0.50100  &       QSO & 3390.8 &  16.8 &   86.6 &   21.7 &    2.6 &  0.6 & $-$0.12 & $-$0.37 & 27.20 &    2.0  \\
2131$-$021 &  3     &   0.55700  &       QSO & 2272.8 &  17.9 &   79.7 &   20.0 &    3.5 &  0.9 & $-$0.17 & $-$0.05 &  ...  &    1.2  \\
2142$-$758 &  1     &   1.13900  &       QSO &  687.9 &   2.2 &   68.4 &   10.5 &    9.9 &  1.5 &    ...  &    0.49 &  ...  &    ...  \\
2155$-$152 &  3     &   0.67200  &       QSO & 2505.5 &  17.1 &   60.5 &   14.7 &    2.4 &  0.6 &    0.42 & $-$0.26 & 27.48 &    4.0  \\
2203$-$188 &  2,3   &   0.61900  &       QSO & 2200.0 &  11.6 &   85.8 &   11.5 &    3.9 &  0.5 &    0.29 &    0.52 & 27.78 &    0.2  \\
2204$-$540 &  1,3   &   1.20600  &       QSO & 1306.6 &   2.2 &   86.1 &    2.0 &    6.6 &  0.2 &    ...  &    0.46 &  ...  &    ...  \\
2206$-$237 &  2,3   &   0.08700  &       QSO &  340.6 &   2.5 &   24.6 &    3.8 &    7.2 &  1.1 &    0.53 &    0.83 &  ...  &    0.2  \\
2210$-$257 &  2,3   &   1.83300  &       QSO &  609.1 &  15.2 &   29.3 &    7.6 &    4.8 &  1.3 &    0.11 &    0.41 & 27.94 &    1.4  \\
2216$-$038 &  3     &   0.90100  &       QSO & 2571.6 &  13.5 &   66.9 &   17.5 &    2.6 &  0.7 &    0.30 & $-$0.41 & 27.62 &    0.7  \\
2223$-$052 &  3     &   1.40400  &       QSO & 8394.5 &  10.8 &  359.7 &   10.8 &    4.3 &  0.1 &    0.39 & $-$0.47 & 28.49 &    4.5  \\
2227$-$088 &  3     &   1.56100  &       QSO & 1570.5 &   8.1 &$<$40.2 &    8.0 & $<$2.6 &  0.5 & $-$0.47 &    0.09 & 27.81 &    1.3  \\
2243$-$123 &  3     &   0.63000  &       QSO & 2043.7 &   9.8 &   48.8 &   16.4 &    2.4 &  0.8 & $-$0.20 &    0.13 & 27.43 &    1.4  \\
2255$-$282 &  2,3   &   0.92600  &       QSO &11226.5 &  20.3 &  379.2 &    6.7 &    3.4 &  0.1 & $-$0.27 & $-$1.40 & 27.55 &    1.0  \\
2311$-$452 &  1,3   &   2.88400  &       QSO &  585.6 &   1.6 &   28.1 &    4.5 &    4.8 &  0.8 &    ...  &    0.70 &  ...  &    ...  \\
2326$-$477 &  1,3   &   1.30600  &       GAL & 1571.0 &   2.8 &   23.1 &    3.6 &    1.5 &  0.2 &    ...  &    0.30 &  ...  &    ...  \\
2329$-$162 &  2,3   &   1.15500  &       QSO & 1964.4 &   8.1 &   99.2 &   19.7 &    5.0 &  1.0 &    0.18 & $-$0.47 & 27.63 &    1.1  \\
2331$-$240 &  2,3   &   0.04770  &       QSO & 1095.7 &   3.5 &   25.8 &    5.8 &    2.4 &  0.5 & $-$0.26 & $-$0.00 &  ...  &    1.0  \\
2333$-$528 &  1,3   &     ...    &       QSO & 1113.9 &   0.9 &   23.3 &    3.3 &    2.1 &  0.3 &    ...  &    0.07 &  ...  &    ...  \\
2345$-$167 &  2,3   &   0.57600  &       QSO & 2747.7 &   5.1 &   46.8 &    5.4 &    1.7 &  0.2 & $-$0.25 &    0.21 & 27.53 &    2.8  \\
2353$-$686 &  1,3   &   1.71600  &       QSO &  785.4 &   0.9 &   16.4 &    3.4 &    2.1 &  0.4 &    ...  &    0.25 &  ...  &    ...  \\
2354$-$117 &  3     &   0.96000  &       QSO & 1197.1 &   7.2 &   54.0 &   12.3 &    4.5 &  1.0 &    0.16 &    0.16 & 27.62 &    1.5  \\
2355$-$534 &  1,3   &   1.00600  &       QSO & 1602.1 &   2.7 &   51.9 &    6.9 &    3.2 &  0.4 &    ...  & $-$0.04 &  ...  &    ...  \\
\hline
\end{tabular}
\end{center}
\end{table*}


\normalsize
\begin{thebibliography}{99}
  \bibitem{} Aller, M.F., Aller, H.D., Hughes, P.A., \& Latimer, G.E. 1999, ApJ, 512, 601
  \bibitem{} Condon, J.J., Cotton, W.D., Greisen, E.W., Yin, Q.F., Perley, R.A.,
  Taylor, G.B., \& Broderick, J.J. 1998, AJ,  115, 1693
  \bibitem{} De Zotti, G., Gruppioni, C., Ciliegi, P., Burigana, C., Danese, L. 1999, NewA, 4, 481
  \bibitem{} Feigelson, E.D., \& Nelson, P.I., 1985, ApJ, 293, 192
  \bibitem{} Isobe, T., \& Feigelson, E.D., 1990, Bull. Amer. Astro. Society, 22, 917
  \bibitem{} Isobe, T., Feigelson, E.D., \& Nelson, P.I., 1986, ApJ, 306, 490
  \bibitem{} Jones, T.W., Rudnick, L., Aller, H.D., {\it et al.} 1985, ApJ, 290, 627
  \bibitem{} Klein, U., Mack, K.-H., Gregorini, L., \& Vigotti, M., 
2003, A\&A, 406, 579    
  \bibitem{} Kogut, A., Spergel, D.N., Barnes, C.,  {\it et al.} 2003, ApJ,
submitted (astro-ph/0302214)
  \bibitem{} Kovac, J.M., Leitch, E.M., Pryke, C., Carlstrom, J.E., Halverson,
N.W., \& Holzapfel, W.L. 2002, Nature, 420, 772
  \bibitem{} K\"uhr, H., Witzel, A., Pauliny-Toth, I.I.K., \& Nauber U. 1981,
A\&AS,  45, 367  
  \bibitem{} Mesa, D., Baccigalupi, C., De Zotti, G., Gregorini, L., Mack, K.-H.,
Vigotti, M., \& Klein, U. 2002, A\&A, 396, 463
  \bibitem{} Nartallo, R., Gear, W.K., Murray, A.G., Robson, E.I., \& Hough, J.H. 1998,
MNRAS, 297, 667
  \bibitem{} Okudaira A., Tabara, H., Kato, T., \& Inoue, M. 1993, PASJ, 45, 153
  \bibitem{} Pentericci, L., Van Reeven, W., Carilli, C.L., R\"ottgering,
  H.J.A.,\& Miley, G.K. 2000, A\&AS, 145, 121
  \bibitem{} Rudnick, L., Jones, T.W., Aller, H.D., {\it et al.}, 1985, ApJ Suppl., 57, 693
  \bibitem{} Saikia, D.J., \& Salter, C.J. 1988,  ARA\&A, 26, 93
  \bibitem{} Sazhin, M.V., \& Korol\"ev, V.A. 1985, Soviet Astronomy Letters, 11, 204
  \bibitem{} Sault, R.J., Teuben, P.J., \& Wright, M.C.H, ``A retrospective
view of Miriad'' in {\it Astronomical Data Analysis Software and
Systems}, ed. R. Shaw, H.E. Payne, J.J.E. Haynes, ASP Conf.
Series, 77, 433$-$43
  \bibitem{} Stickel, M., Meisenheimer, K., \& K\"uhr, H., 1994, A\&AS, 105, 211
\end{thebibliography}
\end{document}